\documentclass[journal]{IEEEtran}

\usepackage{graphicx} 
\usepackage{import}
\usepackage[acronym]{glossaries}
\makeglossaries
\usepackage{cite}
\usepackage{eucal}
\usepackage{amsmath,amssymb,amsfonts,acronym}
\usepackage{multirow}
\usepackage{adjustbox}
\usepackage{fancyhdr}
\usepackage{makecell}
\usepackage{subcaption} 
\usepackage{graphicx}
\usepackage{enumitem}
\usepackage[table,xcdraw]{xcolor}
\usepackage{color}
\usepackage{float}
\usepackage{titlesec}
\usepackage{tikz}
\usetikzlibrary{positioning,shapes,arrows.meta,calc}
\usepackage{amsthm}
\usepackage[normalem]{ulem}

\DeclareMathOperator{\imag}{Im}
\DeclareMathOperator{\re}{Re}

\begin{document}

\title{Harnessing Mutual Coupling in Cell-Free mMIMO}

\author{Erind~Tufa,
        Antonio A.~D'Amico,~\IEEEmembership{Senior~Member,~IEEE,}
        and~Luca~Sanguinetti,~\IEEEmembership{Fellow,~IEEE} \vspace{-0.7cm}
\thanks{The authors are with the University of Pisa, Dipartimento di Ingegneria dell’Informazione, Via Caruso 16, 56122 Pisa, Italy (e-mail: erind.tufa@phd.unipi.it; antonio.damico@unipi.it; luca.sanguinetti@unipi.it).
This work was supported in part by the Smart Networks and Services Joint Undertaking (SNS JU) under the European Union’s Horizon Europe research and innovation program under Grant Agreement No 101192369 (6G-MIRAI), and in part by the project FoReLab (Departments of Excellence), funded by the Italian Ministry of University and Research (MUR).
}}

\maketitle
\thispagestyle{plain}
\pagestyle{plain}

\begin{abstract}
This letter investigates a cell-free massive MIMO system with $L$
 access points (APs), each employing a small fixed-form-factor, uniform linear array of $N$
 potentially closely spaced antennas. Utilizing multiport communication theory, we analyze the impact of mutual coupling on uplink and downlink spectral efficiency (SE) under various impedance-matching strategies. We compare two deployment regimes: ($i$) increasing 
$N$ while $L$ is kept fixed, and ($ii$) increasing 
$N$ while fixing the total antenna count $M = NL$. Numerical results show that, with a full matching network, mutual coupling itself becomes beneficial when an AP serves several user equipments, which is the typical operating regime of user-centric cell-free networks. Furthermore, with full matching, antenna densification can enhance SE or allow for a reduction in the number of APs without performance loss. Since full matching networks remain of manageable size in the small arrays of distributed APs, unlike in large centralized arrays, cell-free networks offer a natural setting for exploiting mutual coupling.
\end{abstract}

\begin{IEEEkeywords}
Mutual coupling, dense arrays, cell-free massive MIMO, multiport communication theory, spectral efficiency.\vspace{-0.4cm}
\end{IEEEkeywords}

\section{Introduction}

The growing demand for high data rates and reliable connectivity has fueled research on advanced multi-antenna technologies. Massive multiple-input multiple-output (mMIMO) systems have emerged as a key enabler \cite{Marzetta2010}, offering substantial gains in spectral efficiency (SE) and link reliability \cite{massiveMIMObook}. Cell-free mMIMO extends this concept by distributing antennas across many APs that jointly serve the user equipments (UEs), eliminating cell boundaries and reducing interference \cite{scalablecellfree}.

In conventional mMIMO deployments, base stations typically employ large arrays with hundreds of antennas and approximately half-wavelength spacing, whereas cell-free systems often rely on small arrays at each AP. This architectural difference facilitates the use of denser antenna configurations, including sub–half-wavelength spacings. In these compact arrays, mutual coupling becomes significant \cite{balanis2016}; although its effects are generally modest in large arrays \cite{damico24holographic}, they may be considerably more pronounced in small arrays. Understanding these effects is therefore essential for assessing the potential of dense antenna deployments in cell-free mMIMO.

Multiport communication theory provides a circuit-theoretic framework to model mutual coupling via impedance matrices \cite{Nossek2010,Nossek2014,damico24holographic,wallace2004,Janaswamy2002}, enabling a physically consistent performance analysis. Previous studies have examined coupling effects on signal-to-noise ratio (SNR) and spectral efficiency, but mainly in single-user or large centralized massive MIMO scenarios \cite{wallace2004,Laas2020}, or at the antenna level, where superdirectivity, matching limitations, and realized gain of compact arrays are well characterized \cite{Deshpande2026parasitic,Kanbaz2024DE,Kanbaz2024howmany,Dovelos2023THz,Nossek2024RIS}.

\begin{figure}[t!]
    \centering
    \includegraphics[angle=90,width=0.9\linewidth]{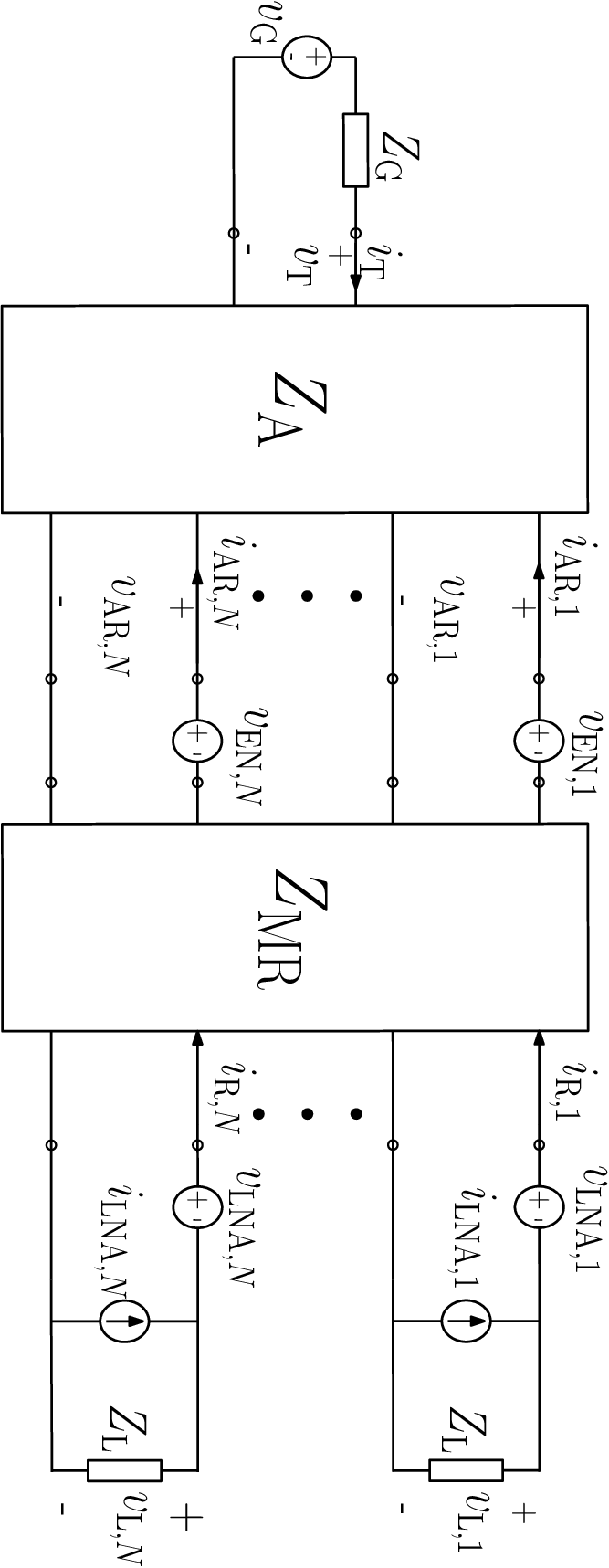}
    \caption{ Circuit model of a single-antenna UE communicating with a multiple-antenna AP.}\centering\vspace{-0.4cm}
    \label{fig:Circuit-model}
\end{figure}

Building on \cite{damico24holographic}, this letter investigates cell-free mMIMO with small, closely spaced uniform linear arrays (ULAs) at each AP. Unlike large base-station arrays, the modest per-AP size (fewer than ten elements) keeps full multiport matching networks (MNs) of manageable size, enabling mutual coupling to enhance performance. We first show that array densification (i.e., increasing the number of antennas within a fixed aperture) provides notable gains, especially in overloaded regimes, highlighting the effectiveness of compact arrays for serving multiple users per AP. 
Then, two deployment scenarios are considered to evaluate the performance trade-offs: ($i$) a fixed number of APs, for which densification results in an increased number of antennas per AP; and ($ii$) a fixed total number of antennas, for which densification leads to a reduced number of APs.
Numerical simulations demonstrate that with a full matching network, densification can increase SE or reduce the number of deployed APs without performance loss, and that these gains stem from mutual coupling itself rather than array gain alone.

\vspace{-0.3cm}
\section{System model}
\vspace{-0.2cm}
We consider a user-centric cell-free mMIMO system with $L$ APs and $K$ single-antenna UEs. Each AP has a ULA of $N$ half-wavelength dipoles, centered at the AP position, with inter-element spacing $d$, giving an array aperture of $\Delta= (N-1)d$.  We consider time division duplexing (TDD) operation with coherence blocks of $\tau_c$ channel uses, divided into $\tau_p$ uplink (UL) pilots, $\tau_u$ UL data, and $\tau_d$ downlink (DL) data. Each AP performs local least-squares (LS) channel estimation \cite{scalablecellfree}. Following \cite[Sec. IV]{scalablecellfree}, we adopt a dynamic cooperation clustering (DCC) approach where each UE is served by a subset of APs. This is captured by diagonal matrices $\mathbf{D}_{kl} \in \mathbb{C}^{N \times N}$, whose entries indicate which antenna elements at AP $l$ are active in serving UE $k$.

\vspace{-0.3cm}

\subsection{Multiport circuit model}
The physical model, based on a circuit-theoretic approach, is shown in Fig. \ref{fig:Circuit-model}, while a complete description of the quantities involved can be found in \cite{damico24holographic}. Specifically, the impedance matrix $\mathbf{Z}_{\rm A}$ quantifies the mutual coupling between antennas and can be written as:
\begin{align}
    \label{Z_A}
    {\bf Z}_{{\rm A}}&= \begin{bmatrix}
    \bf{Z}_{\rm AT}& \bf{0}  \\
    \bf{Z}_{\rm ART} & \bf{Z}_{\rm AR}  \\
    \end{bmatrix}
\end{align}
where $ {\bf Z}_{\rm AT}$ and $ {\bf Z}_{\rm AR}$ depend on the mutual coupling at the transmitter and receiver sides (\textit{intra-array coupling}), respectively, and 
$ {\bf Z}_{\rm ART}$ models the mutual coupling between the receive and transmit arrays (\textit{inter-array coupling}), while the zero block in \eqref{Z_A} follows from the unilateral approximation \cite{damico24holographic}.
The receive matching network is a multiport system described by the impedance matrix ${\bf Z}_{\rm MR} \in \mathbb{C}^{2N\times 2N}$ given by
\begin{equation}
    {\bf Z}_{\rm MR} = \left[\begin{array}{cc} {\bf Z}_{{\rm MR},11} & {\bf Z}_{{\rm MR},12} \\ {\bf Z}_{{\rm MR},21} & {\bf Z}_{{\rm MR},22} \end{array}\right]
\end{equation}
with $\{{\bf Z}_{{\rm MR},ij}\in \mathbb{C}^{N \times N}; i =1,2,  j =1,2\}$. We assume a lossless, reciprocal \cite{Nossek2010}, and noiseless \cite{Nyquist1928} matching network. The receive network ${\bf Z}_{{\rm MR}}$ is designed to maximize the SNR, as in \cite[Eq. (28)]{damico24holographic}:
\begin{align}
    \label{Z_MR}
    \!\!\!{\bf Z}_{{\rm MR}}^\star\!&=\! \begin{bmatrix}
    \!\!\!\!\!\!\mathrm{j}{\imag\{Z_{\rm opt}\}}{\bf I}_\mathrm{N} & \!\!\!\!\!\!\!\!\!\!\!\!\!\!\!\!\!\!\mathrm{j} \sqrt{\re\{Z_{\rm opt}\}}\re\{{\bf Z}_{{\rm AR}}\}^{1/2}  \\
    \mathrm{j} \sqrt{\re\{Z_{\rm opt}\}}\re\{{\bf Z}_{{\rm AR}}\}^{1/2} & \!\!\!\!\!\!\!\!\!\!\!\!\!\!\!\!\!\!-\mathrm{j}{\imag\{{\bf Z}_{\rm AR}\}}  \\
    \end{bmatrix}\!\!\!
\end{align}
where $Z_{\rm opt}$ is defined in \cite[Sec. II.G]{damico24holographic}.
Designing matching networks for large arrays is challenging because the number of reactive components in a full matching network grows quadratically with the number of elements \cite{Nossek2010,Nossek2014,Mezghani2024}. A practical alternative is self-impedance matching (SI) \cite[Sec. III.B]{Warnick2009}, which treats each antenna independently. A simpler option is to omit matching networks and directly connect the RF front-end to the antennas.

\vspace{-0.4cm}

\subsection{Channel model}
In the UL, the channel vector $\mathbf{h}_{kl}^{\rm ul}$ between UE $k$ and AP $l$ is computed as \cite[Eq. (38)]{damico24holographic}:
\begin{equation}
    \mathbf{h}_{kl}^{\rm ul} = \mathbf{Q}^{\rm ul} \mathbf{F}^{\rm ul}_{\rm R} \mathbf{d}_{\text{OC},kl}^{\rm ul}
    \label{eq:channel}
\end{equation}
with  $\mathbf{Q}^{\rm ul}$ and $\mathbf{F}^{\rm ul}_{\rm R}$ being defined in \cite[Eqs. 17 and 15]{damico24holographic}, while $\mathbf{d}_{\text{OC},kl}^{\rm ul}$ represents the open-circuit voltage response at the AP antenna ports.
We assume a line-of-sight scenario where the received field at each antenna can be locally approximated as a plane wave. Since $\Delta = \lambda$, the Fraunhofer distance $2\Delta^2/\lambda = 6$~cm is far below any AP--UE distance, so all UEs are in the far field. In this case, \(\mathbf{d}^{\rm ul}_{\text{OC},kl}\) is given by \cite[Eq. (35)]{damico24holographic}:
\begin{equation}
    \mathbf{d}^{\rm ul}_{\text{OC},kl} = e^{j\phi_{kl}} \boldsymbol{\alpha}(\boldsymbol{\psi}_{kl}, \mathbf{r}_{kl}) \odot \mathbf{a}(\mathbf{r}_{kl})
\end{equation}
where the vectors \(\boldsymbol{\psi}_{kl}\) and \(\mathbf{r}_{kl}\) collect the directions of arrival and distances of the signals from UE \(k\), \(\boldsymbol{\alpha}(\boldsymbol{\psi}_{kl}, \mathbf{r}_{kl})\) is the vector of channel gains, and \(\mathbf{a}(\mathbf{r}_{kl})\) is the array response vector. The phase shift ${\phi_{kl}}\sim \mathcal{U}(0, 2\pi)$ is drawn independently for each UE–AP pair and captures channel randomness. When arbitrary matching networks are used at the AP (e.g. self-impedance matching networks), the DL channel vector $\mathbf{h}_{kl}^{\rm dl}$ between AP $l$ and UE $k$ is obtained as a linear transformation of the UL channel \cite[Sec.~IV-C]{damico24holographic}:
\begin{equation}
    \mathbf{h}_{kl}^{\rm dl} = \mathbf{T}_{\rm ul,dl} \mathbf{h}_{kl}^{\rm ul}
    \label{eq:downlink_channel}
\end{equation}
where $ \mathbf{T}_{\rm ul,dl} = \frac{\alpha_{\rm dl}}{\alpha_{\rm ul}}\mathbf{B}_{\rm dl}^{-{\rm T}/2}\mathbf{A}_{\rm dl,ul}$. The expressions for $\alpha_{\rm ul}$, $\alpha_{\rm dl}$, $\mathbf{B}_{\rm dl}$ and $\mathbf{A}_{\rm dl,ul}$ are given in \cite{damico24holographic}. 
With a full matching network, $\mathbf{T}_{\rm ul,dl}$ reduces to a scaled identity matrix \cite[Eqs.~(58)--(61)]{damico24holographic}.

\vspace{-0.2cm}
\section{Spectral efficiency}

We use $\tau_p$ orthogonal pilots of length $\tau_p \le K$, assigned upon network access. The LS estimate of ${\mathbf{h}}_{kl}^{\rm ul}$ is \cite{demir2021foundations}
\begin{align}
\widehat{\mathbf{h}}_{kl}^{\rm ul} = \mathbf{h}_{kl}^{\rm ul} + \sum\nolimits_{i \in \mathcal{S}_k \setminus \{k\}}\mathbf{h}_{il}^{\rm ul} + \frac{1}{\sqrt{\tau_p \rho}} \mathbf{n}_{l} ^{\rm ul}
\end{align}
where $\rho$ is the UL pilot power and $\mathcal{S}_k$ the set of UEs sharing UE $k$'s pilot. The Gaussian noise $\mathbf{n}_l^{\rm ul}\sim \mathcal{CN}(\mathbf{0}, \mathbf{R}_{\mathbf{n}}^{\prime \rm ul})$ has covariance matrix $\mathbf{R}_{\mathbf{n}}^{\prime \rm ul} = \mathbf{Q}^{\rm ul} \mathbf{U}^{\rm ul} (\mathbf{Q}^{\rm ul})^{\mathrm{H}}$, where $\mathbf{U}^{\rm ul}$ is defined in \cite[Sec. II.F]{damico24holographic}. We call $\widehat{\mathbf{h}}_{k}^{\rm ul}\in \mathbb{C}^{LN}$ the stacked UL channel estimate. From \eqref{eq:downlink_channel}, the corresponding DL estimate is $\widehat{\mathbf{h}}_{k}^{\rm dl} =  (\mathbf{I_{\mathrm{L}}}\otimes \mathbf{T}_{\rm ul,dl}) \widehat{\mathbf{h}}_{k}^{\rm ul}$.

\subsection{Uplink Data Transmission}

To detect the data from UE $k$ in the UL, the central processing unit (CPU) selects a receive combining vector $\mathbf{v}_{k}\in \mathbb{C}^{LN}$.
An achievable SE of UE $k$ is given by \cite[Sec. V]{demir2021foundations}
\begin{align}\label{eq:spectral_efficiency_uplink}
  \text{SE}_k^\text{ul} = \frac{\tau_u}{\tau_c} \log_2(1 + \text{SINR}_k^\text{ul})
\end{align}
with the effective $\text{SINR}_k^\text{ul}$ defined as
\begin{equation}
\label{eq:09}
\resizebox{0.91\columnwidth}{!}{$%
  \frac{\left| \mathbb{E} \left\{ \mathbf{v}_{k}^\mathrm{H} \mathbf{D}_k \mathbf{h}_{k}^{\rm ul} \right\} \right|^2}{\sum_{i=1}^K \mathbb{E} \left\{ \left| \mathbf{v}_{k}^\mathrm{H} \mathbf{D}_k \mathbf{h}_{i}^{\rm ul} \right|^2 \right\} - \left| \mathbb{E} \left\{ \mathbf{v}_{k}^\mathrm{H} \mathbf{D}_k \mathbf{h}_{k}^{\rm ul} \right\} \right|^2 + \frac{1}{p^\text{ul}} \mathbb{E} \left\{ \mathbf{v}_k^\mathrm{H} \mathbf{D}_k\mathbf{R}_{\mathbf{n}}^\text{ul}\mathbf{D}_k\mathbf{v}_k \right\}}
$}
\end{equation}

where $p^\text{ul}$ is the UL symbol power of one UE, related to the physical transmit power $P_T$ of Table \ref{tab:parameters} as in \cite[Eq.~(42)]{damico24holographic} and $ {\mathbf D}_k = {\rm blkdiag} \left( {\mathbf D}_{k1}, \dots ,{\mathbf D}_{kL}\right)$, where $\left( {\mathbf D}_{k1}, \dots ,{\mathbf D}_{kL}\right)$ are the matrices characterizing the DCC policy \cite{ demir2021foundations}, and $\mathbf{R}_{\mathbf{n}}^{\rm ul} = \mathbf{I_{\mathrm{L}}} \otimes \mathbf{R}_{\mathbf{n}}^{\prime \rm ul} $. Equal power allocation is assumed in uplink and downlink. The expectation in \eqref{eq:09} is over all randomness. While \eqref{eq:spectral_efficiency_uplink} holds for any combiner, here we adopt the minimum mean square error (MMSE) combiner \cite{demir2021foundations}.

\vspace{-0.4cm}
\subsection{Downlink Data Transmission}
In the DL, the CPU coordinates the APs’ transmissions to the UEs. As in the UL, an achievable SE for UE $k$ is:
\begin{align}
  \text{SE}_k^\text{dl} = \frac{\tau_d}{\tau_c}\log_2(1 + \text{SINR}_k^\text{dl})
\end{align}
with the effective $\text{SINR}_k^\text{dl}$ defined as

\begin{equation}
\resizebox{0.89\columnwidth}{!}{$\displaystyle
\frac{p_k^\text{dl} \left| \mathbb{E} \left\{ \left( \mathbf{h}_{k}^{\rm dl} \right)^\mathrm{T} \mathbf{D}_k \tilde{\mathbf{w}}_{k} \right\} \right|^2}{\sum\limits_{i=1}^K p_i^\text{dl} \mathbb{E} \left\{ \left| \left( \mathbf{h}_{k}^{\rm dl} \right)^\mathrm{T} \mathbf{D}_i \tilde{\mathbf{w}}_{i} \right|^2 \right\} - p_k^\text{dl} \left| \mathbb{E} \left\{ \left( \mathbf{h}_{k}^{\rm dl} \right)^\mathrm{T} \mathbf{D}_k \tilde{\mathbf{w}}_{k} \right\} \right|^2 + \sigma^2_{\rm dl}}
$}
\end{equation}
where $p_{k}^\text{dl}$ is the DL transmit power for UE $k$ and $\tilde{\mathbf{w}}_{k}\in \mathbb{C}^{LN}$ is its unit-norm precoder. DL SE depends on all UEs’ precoders, making joint optimization intractable. While UL–DL duality is a common workaround \cite[Theorem 6.2]{demir2021foundations}, it does not hold here due to non-scalar channel transformations and different noise statistics in UL and DL.
We thus adopt a heuristic approach: DL channel estimates are obtained by transforming the UL estimates by \eqref{eq:downlink_channel} and used in an MMSE-type precoder, whose performance will depend on the UL estimation accuracy and the linear transformation $\mathbf{T}_{\rm ul,dl}$, as shown in Section~V-B.

\begin{figure}[t!]
    \centering\centering\vspace{-1cm}
    \includegraphics[width=0.8\linewidth]{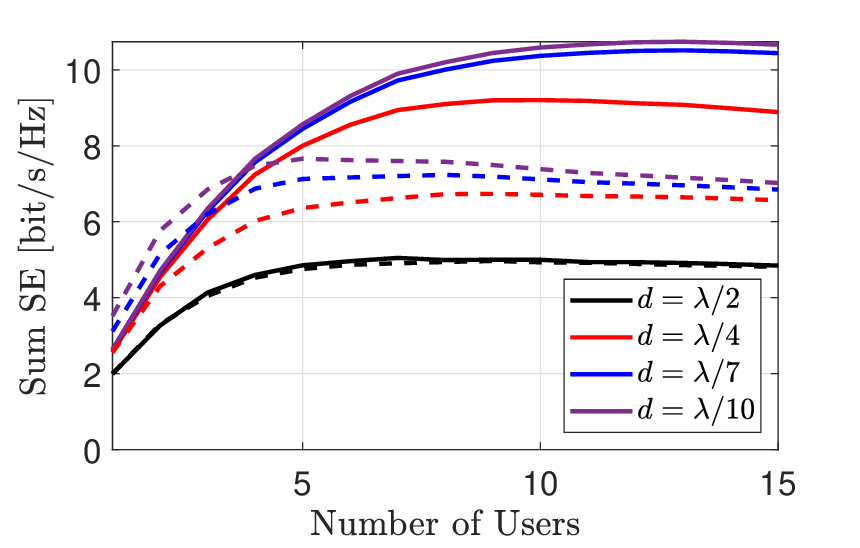}
    \caption{Sum SE as a function of the number of served UEs, with (solid) and without (dashed) mutual coupling, for an MMSE combiner and a full matching network, at inter-element spacings $d \in \{\lambda/2, \lambda/4, \lambda/7, \lambda/10\}$.}
    \label{fig:sumSE_overl}\centering\vspace{-0.3cm}
\end{figure}


\section{Is It All Array Gain? The Role of Mutual Coupling}
Packing more elements into a fixed aperture inevitably increases array gain, but the role of mutual coupling is not obvious. To isolate it, we compare the coupled array against a coupling-free benchmark obtained by reducing $\mathbf{Z}_{\rm AR}$ to its diagonal, i.e., setting the mutual terms to zero. Such an array cannot be realized and serves only as a reference. Everything else is unchanged (element positions, aperture, UE distribution, combiner, noise model) and the MN is re-optimized for each configuration. We consider a single AP with a fixed aperture $\Delta =  \lambda$, where $\lambda$ denotes the wavelength at $10~ \text{GHz}$. The AP serves up to $K = 15$ UEs randomly distributed over a plane, with distances from the array center ranging within $[10, 150]$~m and with azimuth angles spanning $[-\pi/2, \pi/2]$. We use an MMSE combiner and a full MN, as given in~\eqref{Z_MR}.
Fig.~\ref{fig:sumSE_overl} shows the sum spectral efficiency as a function of the number of served UEs for different antenna spacings.
At $d=\lambda/2$ the two configurations coincide at every load, while at the densest spacings coupling is slightly detrimental for $K \le 4$. Mutual coupling becomes beneficial at higher loading levels: the gain grows with $K$ and saturates at $K \approx 9$ for $d=\lambda/4$ and $K \approx 12$ for the densest spacings. The coupling-free curves saturate beyond $K = 5$, confirming that additional antennas alone do not translate into greater user separability. 
Cell-free deployments are therefore a
particularly favorable setting: in a scalable user-centric network, each
AP serves up to \(\tau_p\) UEs \cite{scalablecellfree}, and can thus naturally operate in
this regime, while the small per-AP arrays keep full matching networks
of manageable size.

\vspace{-0.2cm}
\section{Numerical Results}

To assess the performance of closely-spaced ULAs in user-centric cell-free mMIMO networks, we consider two deployment scenarios that highlight different aspects of densification:

\begin{enumerate}[leftmargin=*,topsep=2pt,itemsep=1pt]
    \item \textbf{Fixed number of APs} ($L$): The number of APs remains constant while the number of antenna elements per AP, $N$, is varied. Densification (i.e., reducing $d$) allows more antennas  within a fixed aperture $\Delta$, thereby increasing both $N$ and the total number of antennas in the network, $M = LN$. This scenario emphasizes the impact of exploiting mutual coupling within increasingly dense antenna arrays.
    
    \item \textbf{Fixed total number of antennas} ($M$): The total number of antennas across the network is fixed. As we increase $N$ through densification, the number of APs $L = M/N$ must decrease accordingly. This scenario illustrates the fundamental trade-off between ($i$) maintaining spatial diversity and coverage through distributed deployment of multiple APs, and ($ii$) exploiting mutual coupling effects within fewer but denser arrays.
\end{enumerate}
The system parameters are summarized in Table~\ref{tab:parameters}.

\begin{table}[t]
\centering\centering\vspace{-1cm}
\caption{System Parameters}
\label{tab:parameters}
\begin{tabular}{lc}
\hline
\textbf{Parameter} & \textbf{Value} \\
\hline
Carrier frequency, $f_c$ & 10 GHz \\
Bandwidth, $B$ & 20 MHz \\
Coverage area & $200 \times 200$ m$^2$ \\
AP height & 10 m \\
Number of UEs, $K$ & 15 \\
Number of pilots, $\tau_p$ & 10 \\
UE transmit power, $P_T$ & $-15$ dBm \\
Array aperture, $\Delta$ & $\lambda$ \\
Inter-element spacing, $d$ & \{$\lambda/2, \lambda/3, \lambda/4, \lambda/7$\}\\
\hline
\hline
\textbf{Scenario 1: Fixed number of APs} & \textbf{Value} \\
\hline
Number of APs, $L$ & 40 \\
Total antennas, $M$ & \{120, 160, 200, 320\}\\
\hline
\textbf{Scenario 2: Fixed total antennas} & \textbf{Value} \\
\hline
Number of APs, $L$ & \{40, 30, 24, 15\} \\
Total antennas, $M$ & 120 \\
\hline
\end{tabular}
\end{table}

\begin{figure}[t!]
    \centering\vspace{-0.3cm}
    \includegraphics[width=0.8\linewidth]{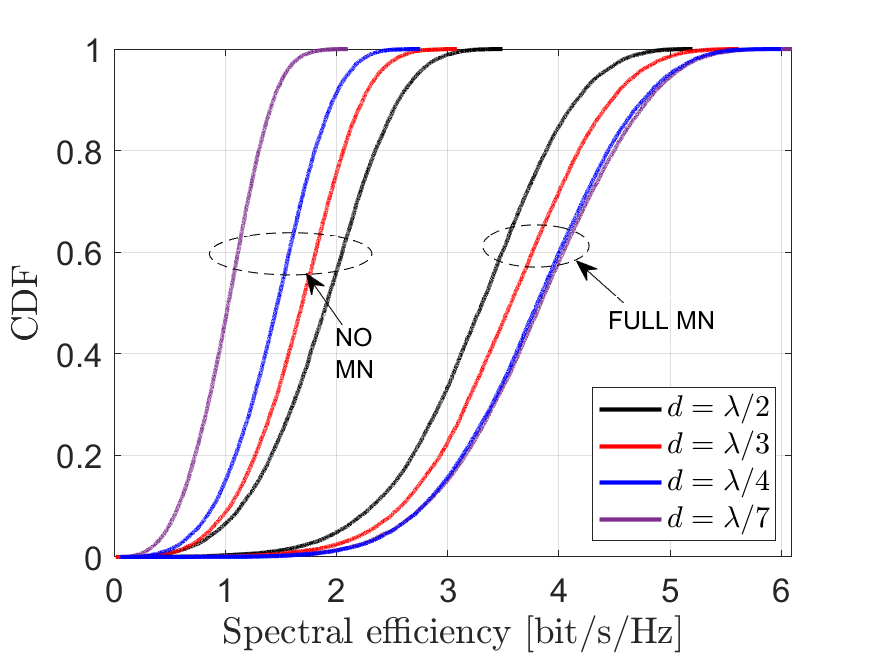}
    \caption{CDF of per-UE UL spectral efficiency for Scenario 1 (fixed number of APs, $L=40$) with different matching network strategies and inter-element spacings $d \in \{\lambda/2, \lambda/3, \lambda/4,\lambda/7\}$.}\centering\vspace{-0.3cm}
    \label{fig:scenario1}
\end{figure}

\vspace{-0.3cm}
\subsection{Uplink Spectral Efficiency}

Fig.~\ref{fig:scenario1} shows the cumulative distribution function (CDF) of the per-UE SE for Scenario 1, where the number of APs is fixed at $L=40$ and the inter-element spacing varies as in Table~\ref{tab:parameters}. Without a matching network, mutual coupling progressively degrades performance as spacing decreases: $d=\lambda/3$ performs worse than $d=\lambda/2$, $d=\lambda/4$ performs worse still, and $d=\lambda/7$ yields the lowest SE. This indicates that, without impedance matching, the adverse effects of mutual coupling outweigh the array-gain benefits of adding antennas per AP.
With a full matching network, the trend reverses. The densest deployments, $d=\lambda/4$ and $d=\lambda/7$, achieve the highest and nearly identical SE, showing that proper matching enables the system to exploit mutual coupling. The negligible difference between these two spacings indicates a saturation point, beyond which further densification provides no additional gain.
The self-impedance matching network (not shown) yields intermediate performance. While $d=\lambda/3$ and $d=\lambda/4$ still outperform the conventional spacing, the gains are smaller than with full matching. For \(d=\lambda/7\), performance falls below that of the conventional array because SI MN neglects off-diagonal coupling terms, causing an increasing impedance mismatch.

\begin{figure}[t!]
    \centering\centering\vspace{-1cm}
    \includegraphics[width=0.8\linewidth]{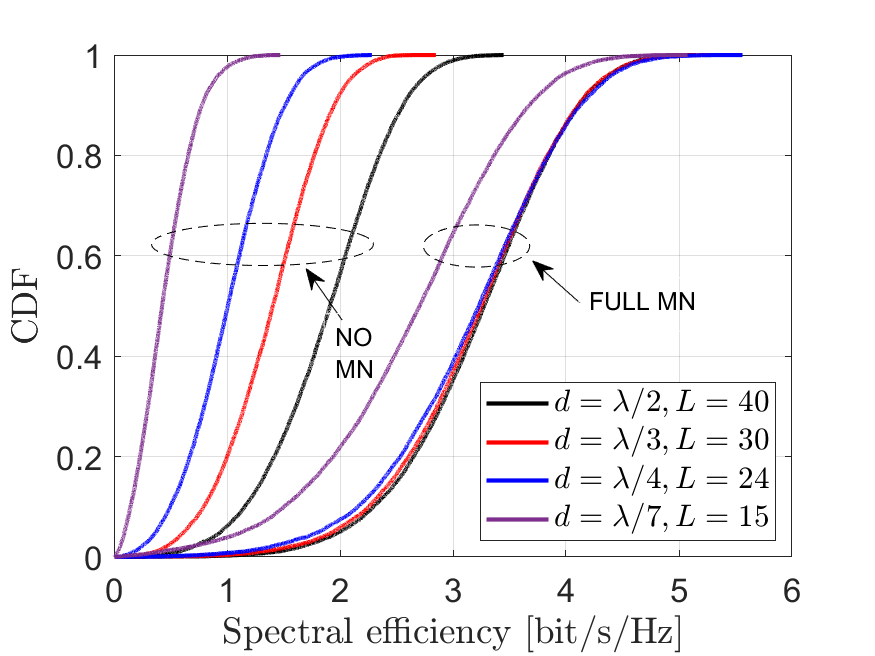}
    \caption{CDF of per-UE UL spectral efficiency for Scenario 2 (fixed total number of antennas, $M=120$) with different matching network strategies and inter-element spacings $d \in \{\lambda/2, \lambda/3, \lambda/4,\lambda/7\}$.}\centering\vspace{-0.3cm}
    \label{fig:scenario2}
\end{figure}

Fig.~\ref{fig:scenario2} shows the CDF of per-UE SE for Scenario~2 ($M=120$ fixed, fewer APs as $d$ decreases). Without a matching network, performance degrades even further, since the unmitigated coupling is now compounded by the reduced spatial diversity. With a full MN, the spacings $d\in\{\lambda/2,\lambda/3,\lambda/4\}$ achieve comparable SE, demonstrating that proper impedance matching allows the number of APs to be reduced without sacrificing performance. At $d=\lambda/7$, however, densification reaches a practical limit: reducing to $L=15$ APs causes insufficient spatial coverage that the denser per-AP arrays cannot compensate, even under full MN. SI MN (not plotted) outperforms the unmatched case but follows the same trend: performance degrades progressively with densification, with $d=\lambda/4$ and $d=\lambda/7$ falling below the conventional spacing.

\vspace{-0.25cm}
\subsection{Downlink Spectral Efficiency}
The UL analysis showed that a full matching network is essential to exploit mutual coupling and that densification beyond $d = \lambda/4$ provides marginal or no advantages. We now examine whether these trends carry over to the DL, where imperfect channel state information (CSI) and the structure of the UL-to-DL channel transformation introduce additional factors.

Under perfect CSI, all MN configurations achieve identical DL performance, which coincides with the full MN UL configuration; the DL power constraint is imposed at the antenna ports and not at the generator output, so all strategies operate under the same effective transmit conditions.

\begin{figure}[t!]
    \centering\vspace{-1cm}
    \includegraphics[width=0.8\linewidth]{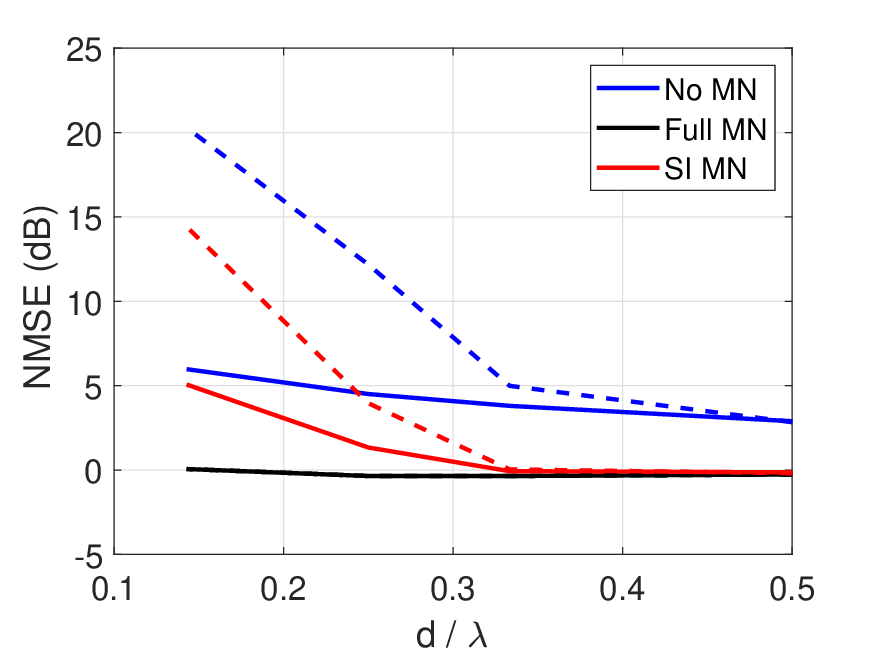}
    \caption{Average NMSE (dB) of the channel estimation as a function of antenna spacing for  different matching strategies. Solid lines represent UL estimation, while dashed lines represent DL estimation.}
    \label{fig:NMSE}\centering\vspace{-0.3cm}
\end{figure}

\begin{figure}[t!]
    \centering\centering
    \includegraphics[width=0.8\linewidth]{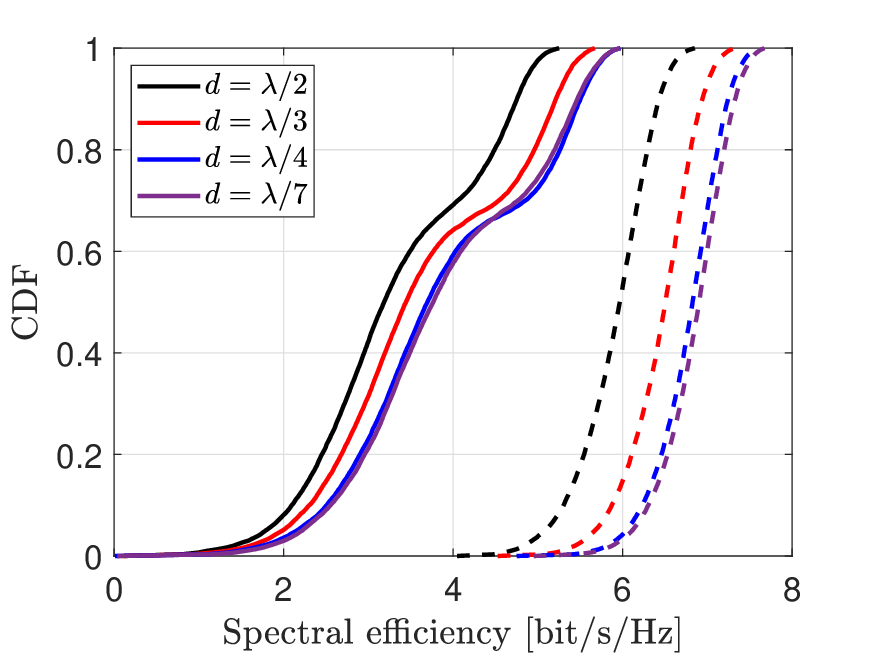}
    \caption{CDF of the per-UE DL spectral efficiency for Scenario 1 under full MN and antenna spacings $d \in \{\lambda/2, \lambda/3, \lambda/4, \lambda/7\}$. Solid lines represent the CDF with imperfect CSI, while dashed lines correspond to perfect CSI.}
    \label{fig:DL_SE_Scenario_1}\vspace{-0.3cm}
\end{figure}

The picture changes under imperfect CSI. As shown in Fig.~\ref{fig:NMSE}, the UL channel estimation accuracy degrades with densification under SI MN and No MN, whereas full MN maintains consistently low normalized mean square error (NMSE) across all spacings. 
In addition, the transformation $\mathbf{T}_{\rm ul,dl}$ introduces a second, DL-specific impairment. Under full MN, $\mathbf{T}_{\rm ul,dl}$ is a scaled identity, so estimation errors are transferred from UL to DL without amplification and the UL and DL NMSE curves coincide. In contrast, under SI MN and No MN, the conditioning of $\mathbf{T}_{\rm ul,dl}$ worsens as $d$ decreases, amplifying UL estimation errors and causing a widening gap between UL and DL NMSE. The combination of degraded UL estimation and error amplification explains the severe DL performance loss at small spacings. Since only full MN maintains acceptable performance across all spacings, the following DL analysis focuses on this case.
Fig.~\ref{fig:DL_SE_Scenario_1} shows the per-UE DL SE for Scenario 1 under full MN. As in the UL, $d=\lambda/4$ achieves the highest SE, while reducing the spacing to $d=\lambda/7$ yields no additional gain, indicating saturation. With imperfect CSI, the same ordering is preserved and the SE loss remains moderate and nearly uniform across spacings, consistent with the flat NMSE profile of the full MN.

\begin{figure}[t!]
    \centering\centering\vspace{-0.9cm}
    \includegraphics[width=0.8\linewidth]{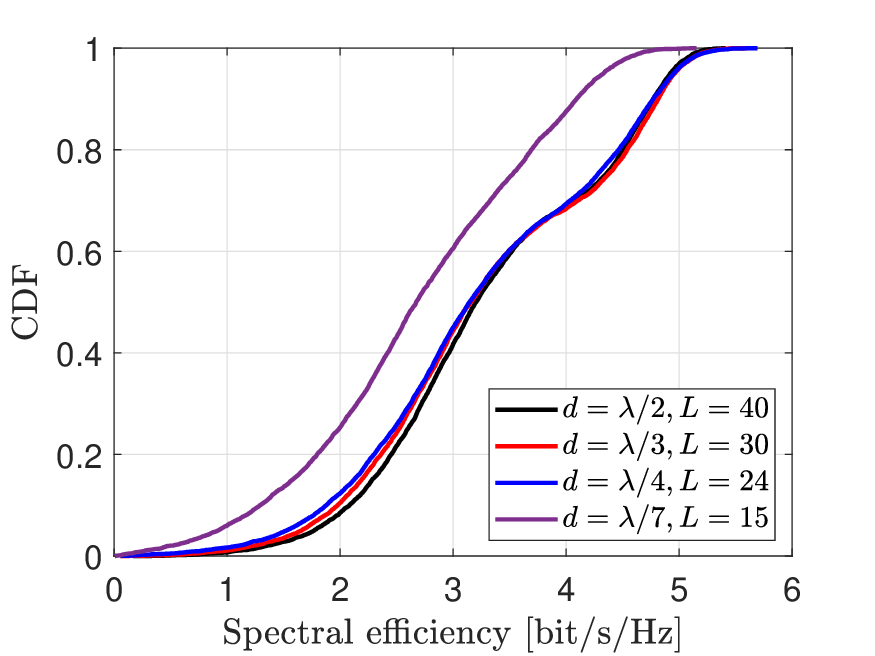}
    \caption{CDF of per-UE DL spectral efficiency for Scenario 2 (fixed total number of antennas, $M=120$) with full MN and inter-element spacings $d \in \{\lambda/2, \lambda/3, \lambda/4,\lambda/7\}$.}
    \label{fig:DL_SE_Scenario_2}\vspace{-0.3cm}
\end{figure}

Fig.~\ref{fig:DL_SE_Scenario_2} presents the per-UE DL SE for Scenario 2 under full MN. The results are consistent with the UL analysis: $d \in \{\lambda/2, \lambda/3, \lambda/4\}$  achieve comparable performance, confirming that proper impedance matching allows the number of APs to be reduced without sacrificing SE. At $d = \lambda/7$,  however, performance degrades noticeably, for the same reason as in the UL. The close agreement between the UL and DL results for the full MN, already anticipated by the NMSE curves in Fig.~\ref{fig:NMSE}, reinforces that, when $\mathbf{T}_{\rm ul,dl}$ reduces to a scaled identity, the DL channel estimation inherits the quality of the UL estimate without amplification, and the precoder performs reliably.



\vspace{-0.25cm}

\subsection{Not just array gain: benefits of mutual coupling}

\begin{figure}[t!]
    \centering\centering\vspace{-0.9cm}
    \includegraphics[width=0.8\linewidth]{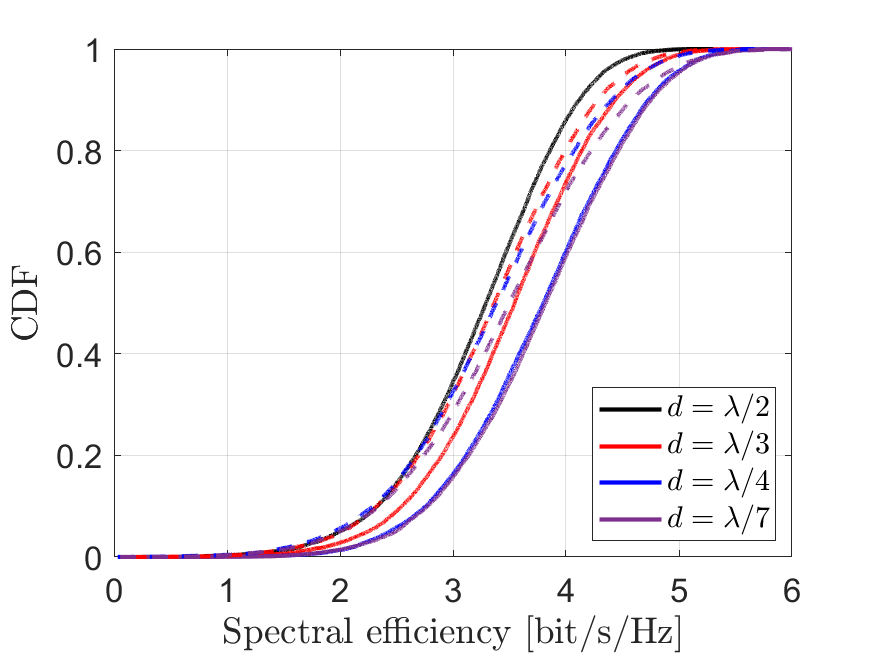}
    \caption{CDF of per-UE UL spectral efficiency for 
    Scenario~1 under full MN, comparing deployments with mutual 
    coupling (solid) and without mutual coupling (dashed) for 
    inter-element spacings $d \in \{\lambda/2, \lambda/3, 
    \lambda/4, \lambda/7\}$.}
    \label{fig:Scenario1_wo_coupling}\vspace{-0.3cm}
\end{figure}

Fig.~\ref{fig:Scenario1_wo_coupling} revisits Scenario~1 under full MN, comparing the system with and without mutual coupling (solid and dashed lines, respectively) across all spacings, mirroring the 
single-AP experiment of Section~IV. As densification increases, coupling becomes progressively beneficial: the gain grows from $d=\lambda/3$ to $d=\lambda/4$, confirming that the array-gain baseline alone 
does not account for the SE improvements observed in 
Fig.~\ref{fig:scenario1}. At $d=\lambda/7$ the gap narrows relative to $d=\lambda/4$, consistent with the saturation already noted in the UL analysis. These results confirm that the SE gains from densification in the cell-free setting arise from mutual coupling itself, not merely from the increased array gain, in line with the overloaded single-AP regime analyzed in Section~IV.

\vspace{-0.28cm}
\section{Conclusion}
This letter studied the impact of mutual coupling on UL and DL spectral efficiency in cell-free massive MIMO, modelling closely spaced ULAs at each AP within the multiport communication framework. Two deployment scenarios were considered: $(1)$ a fixed number of APs, where densification increases the number of antennas per AP and the total number of antennas in the network; and $(2)$ a fixed total number of antennas, where densification reduces the number of APs. The analysis established three principal findings. First, a full matching network is necessary to translate densification into SE gains: without it, mutual coupling uniformly degrades performance regardless of spacing, overwhelming the array gain from additional antennas. Second, with full matching, densification to $d = \lambda/4$ improves SE in Scenario~1 and preserves it with fewer APs in Scenario~2, but further reduction to $d = \lambda/7$ yields no additional gain in Scenario~1 and causes performance loss in Scenario~2. This gain comes from mutual coupling itself, not from the larger element count, and appears when an AP serves several UEs, which is the regime in which the APs of a user-centric cell-free network naturally operate. Third, the DL under full matching reproduces the UL behaviour faithfully: UL estimation errors are transferred to the DL without amplification, and the precoder performs reliably across all retained spacings. Taken together, these results show that cell-free networks are a natural setting for exploiting mutual coupling: the small per-AP arrays keep full multiport matching networks manageable, and the APs operate in the regime where coupling is beneficial, enabling antenna densification to either enhance spectral efficiency or reduce infrastructure deployment without performance penalty. A complete characterization of practical loss mechanisms (conductor and dielectric losses, radiation efficiency, and matching-network insertion loss) is an important direction for future work.
\vspace{-0.25cm}

\vspace{-0.25cm}
\bibliographystyle{IEEEtran}
\bibliography{biblio_v2}

@ARTICLE{Nossek2024RIS,
  author={Nossek, Josef A. and others},
  journal={IEEE Wireless Commun. Lett.}, 
  title={Physically Consistent Modeling of Wireless Links With Reconfigurable Intelligent Surfaces Using Multiport Network Analysis}, 
  year={2024},
  volume={13},
  number={8},
  pages={2240-2244},
  doi={10.1109/LWC.2024.3410528}}

@ARTICLE{Dovelos2023THz,
  author  = {Dovelos, Konstantinos and others},
  journal = {IEEE Trans. Commun.},
  title={Superdirective Antenna Pairs for Energy-Efficient Terahertz Massive MIMO}, 
  year={2023},
  volume={71},
  number={12},
  pages={7316-7332},
  doi={10.1109/TCOMM.2023.3311449}}

@INPROCEEDINGS{Kanbaz2024howmany,
  author    = {Kanbaz, Ihsan and others},
  booktitle = {Proc. IEEE WCNC},
  title     = {Super-Directive Antenna Arrays: How Many Elements Do We Need?},
  year      = {2024},
  month     = apr,
  pages     = {1--6},
  doi       = {10.1109/WCNC57260.2024.10570713}}

@INPROCEEDINGS{Kanbaz2024DE,
  author    = {Kanbaz, Ihsan and others},
  booktitle = {Proc. EuCAP},
  title={Optimization of Super-Directive Linear Arrays with Differential Evolution for High Realized Gain}, 
  year={2024},
  pages={1-5},
  doi={10.23919/EuCAP60739.2024.10501011}}

@ARTICLE{Deshpande2026parasitic,
  author={Deshpande, Nitish Vikas and others},
  journal = {IEEE Trans. Wireless Commun.},
  title={Beamforming With Hybrid Reconfigurable Parasitic Antenna Arrays}, 
  year={2026},
  volume={25},
  pages={7048-7064},
  doi={10.1109/TWC.2025.3628753}}

@ARTICLE{demir2021foundations,
   title={Foundations of User-Centric Cell-Free Massive {MIMO}},
   volume={14},
   ISSN={1932-8354},
   doi ={10.1561/2000000109},
   number={3-4},
   journal={Found. Trends Signal Process.},
   publisher={Emerald},
   author  = {Demir, {\"O}zlem Tugfe and Bj{\"o}rnson, Emil and Sanguinetti, Luca},
   year={2021},
   month=Jan, pages={162–472}}

@ARTICLE{Warnick2009,
  author  = {Warnick, Karl F. and others},
  journal = {IEEE Trans. Antennas Propag.},
  title   = {Minimizing the Noise Penalty Due to Mutual Coupling for a Receiving Array},
  volume  = {57},
  number  = {6},
  pages   = {1634--1644},
  year    = {2009},
  doi     = {10.1109/TAP.2009.2019898}}

@ARTICLE{Mezghani2024,
  author  = {Mezghani, Amine and others},
  journal = {IEEE BITS Inf. Theory Mag.},
  title={Reincorporating Circuit Theory Into Information Theory}, 
  year={2024},
  volume={4},
  number={1},
  pages={40-58},
  doi={10.1109/MBITS.2023.3346329}}

@ARTICLE{Janaswamy2002,
  author  = {Janaswamy, R.},
  journal = {IEEE Antennas Wireless Propag. Lett.},
  title   = {Effect of Element Mutual Coupling on the Capacity of Fixed Length Linear Arrays},
  volume  = {1},
  pages   = {157--160},
  year    = {2002},
  doi     = {10.1109/LAWP.2002.807570}}

@ARTICLE{Marzetta2010,
  author  = {Marzetta, Thomas L.},
  journal = {IEEE Trans. Wireless Commun.},
  title   = {Noncooperative Cellular Wireless with Unlimited Numbers of Base Station Antennas},
  volume  = {9},
  number  = {11},
  pages   = {3590--3600},
  year    = {2010},
  doi     = {10.1109/TWC.2010.092810.091092}}

@ARTICLE{Laas2020,
  author={Laas, Tobias and others},
  journal={IEEE Trans. Wireless Commun}, 
  title={On Reciprocity in Physically Consistent {TDD} Systems With Coupled Antennas}, 
  year={2020},
  volume={19},
  number={10},
  pages={6440-6453},
  doi={10.1109/TWC.2020.3003414}}

@ARTICLE{wallace2004,
  author  = {Wallace, J. W. and Jensen, M. A.},
  journal = {IEEE Trans. Wireless Commun.},
  title   = {Mutual Coupling in {MIMO} Wireless Systems: A Rigorous Network Theory Analysis},
  volume  = {3},
  number  = {4},
  pages   = {1317--1325},
  year    = {2004},
  doi     = {10.1109/TWC.2004.830854}}

@BOOK{balanis2016,
  author    = {Balanis, Constantine A.},
  title     = {Antenna Theory: Analysis and Design},
  edition   = {4},
  year      = {2016},
  publisher={John Wiley \& Sons},
  address={Hoboken, NJ}}

@ARTICLE{massiveMIMObook,
  author  = {Bj{\"o}rnson, Emil and Hoydis, Jakob and Sanguinetti, Luca},
  journal = {Found. Trends Signal Process.},
  title   = {Massive {MIMO} Networks: Spectral, Energy, and Hardware Efficiency},
  volume  = {11},
  number  = {3--4},
  pages   = {154--655},
  year    = {2017},
  doi     = {10.1561/2000000093}}

@ARTICLE{scalablecellfree,
  author  = {Bj{\"o}rnson, Emil and Sanguinetti, Luca},
  journal = {IEEE Trans. Commun.},
  title   = {Scalable Cell-Free Massive {MIMO} Systems},
  volume  = {68},
  number  = {7},
  pages   = {4247--4261},
  year    = {2020},
  doi     = {10.1109/TCOMM.2020.2987311}}

@ARTICLE{Nyquist1928,
  author  = {Nyquist, H.},
  journal = {Phys. Rev.},
  title   = {Thermal Agitation of Electric Charge in Conductors},
  volume  = {32},
  number  = {1},
  pages   = {110--113},
  year    = {1928},
  doi     = {10.1103/PhysRev.32.110}}

@ARTICLE{Nossek2014,
  author  = {Ivrla{\v{c}}, Michel T. and Nossek, Josef A.},
  journal = {IEEE Circuits Syst. Mag.},
  title   = {The Multiport Communication Theory},
  volume  = {14},
  number  = {3},
  pages   = {27--44},
  year    = {2014},
  doi     = {10.1109/MCAS.2014.2333618}}

@ARTICLE{Nossek2010,
  author  = {Ivrla{\v{c}}, Michel T. and Nossek, Josef A.},
  journal = {IEEE Trans. Circuits Syst. I, Reg. Papers},
  title   = {Toward a Circuit Theory of Communication},
  volume  = {57},
  number  = {7},
  pages   = {1663--1683},
  year    = {2010},
  doi     = {10.1109/TCSI.2010.2043994}}

@ARTICLE{damico24holographic,
  author  = {D'Amico, Antonio Alberto and Sanguinetti, Luca},
  journal = {IEEE Trans. Wireless Commun.},
  title   = {Holographic {MIMO} Communications: What is the Benefit of Closely Spaced Antennas?},
  volume  = {23},
  number  = {10},
  pages   = {13826--13840},
  year    = {2024},
  doi     = {10.1109/TWC.2024.3405199}}

\end{document}